\documentclass[prb,nobalancelastpage,twocolumn,superscriptaddress,showpacs,nofootinbib]{revtex4}

\usepackage{color,amsthm,amsmath,amsfonts,graphicx,bm,amssymb}
\usepackage{epstopdf}
\usepackage{ulem}

\newcommand{\ket}[1]{|#1\rangle}
\newcommand{\bra}[1]{\langle #1|}
\newcommand{\tr}{\text{tr}}
\newcommand{\Ham}{\mathcal H}

\begin{document}

\title{Out-of-equilibrium dynamics and thermalization of string order} 

\author{Leonardo Mazza}
\address{NEST, Scuola Normale Superiore \& Istituto Nanoscienze-CNR, I-56126 Pisa, Italy}

\author{Davide Rossini}
\address{NEST, Scuola Normale Superiore \& Istituto Nanoscienze-CNR, I-56126 Pisa, Italy}

\author{Manuel Endres}
\address{Max-Planck-Institut f\"ur Quantenoptik, D-85748 Garching, Germany}

\author{Rosario Fazio}
\address{NEST, Scuola Normale Superiore \& Istituto Nanoscienze-CNR, I-56126 Pisa, Italy}


\begin{abstract}
  We investigate the equilibration dynamics of string order in one-dimensional quantum systems.
  After initializing a spin-1 chain in the Haldane phase, 
  the time evolution of non-local correlations 
  following a sudden quench is studied by means of matrix-product-state-based algorithms. 
  Thermalization occurs only for scales up to a horizon growing at a well defined speed,
  due to the finite maximal velocity at which string correlations can propagate, 
  related to a Lieb-Robinson bound.
  The persistence of string ordering at finite times is non-trivially related to symmetries 
  of the quenched Hamiltonian.
  A qualitatively similar behavior is found for the string order of the Mott insulating phase
  in the Bose-Hubbard chain. This paves the way towards an experimental testing
  of our results in present cold-atom setups.
\end{abstract}

\maketitle


{\it Introduction.}---Characterizing the long-time properties of many-body systems 
evolving according to a unitary time evolution is a long-standing and fascinating 
task.~\cite{vonNeumann29,Deutsch_1991,Srednicki_1994,chaos_ergod} 
Because of special physical importance,
the equilibration of Landau order parameters 
and of two-point correlators has deserved almost unique attention. 
It is by now assessed, however, that not every quantum phase of matter is amenable 
for a description in terms of local observables. The peculiarity of topological order 
only emerges via the study of non-local operators.~\cite{wen, Haldane1983, DallaTorre2006}

Triggered by impressive experimental progresses on ultracold atomic gases, where both 
the initial state and the Hamiltonian dynamics can be engineered with unprecedented 
accuracy,~\cite{Kinoshita_2006, Trotzky2012, Gring_2012, Cheneau_2012, Fukuhara2013a, Langen2013, Fukuhara2013b} 
last years have witnessed a burst in the theoretical understanding of the equilibration 
of closed systems.~\cite{review1, review2, review3} 
When the dynamics is governed by a non-integrable Hamiltonian, 
such that the energy is the only non-trivial conserved quantity, the features 
of a canonical ensemble are expected to emerge \textit{locally} in the steady 
state.~\cite{Rigol_2008, Moeckel_2008, Roux_2009, Biroli_2010, Pal_2010, Banuls2011, Canovi_2011, Brandino_2012, Carleo_2012} 
The occurrence of thermalization for \textit{non-local} order, instead, 
cannot be predicted via this picture, and stands as an intriguing open problem.
The literature on this topic mainly focused on the equilibration 
dynamics of entropy measures for topological 
order.~\cite{Tsomokos_2009, Rahmani_2010, Halasz2013, Patel_2013} 

Here we study the time evolution after a global quantum quench of a 
string operator for a one-dimensional system. 
Since the string can extend over a macroscopic part of the chain,
its dynamics does not simply fit into the picture of a system acting as the bath of one of its parts.
We start discussing the string order (SO) in a spin-1 chain.~\cite{Haldane1983} 
After quenching the system within or outside the Haldane phase, 
string correlations reveal thermalization on short lengths but the finite velocity 
at which they propagate, related to a Lieb-Robinson (LR) bound, 
prevents this to occur for longer lengths.
Moreover, a symmetry analysis reveals the existence of Hamiltonians destroying the SO abruptly.
The possibility to experimentally access string operators with cold atoms~\cite{Endres_2011} 
allows a direct test of our predictions.
To this aim we conclude by analyzing quenches within the Mott insulating phase 
in the Bose-Hubbard (BH) chain, where this type of experiment 
is most likely to be performed.

\begin{figure}[t]
  \includegraphics[width=\columnwidth]{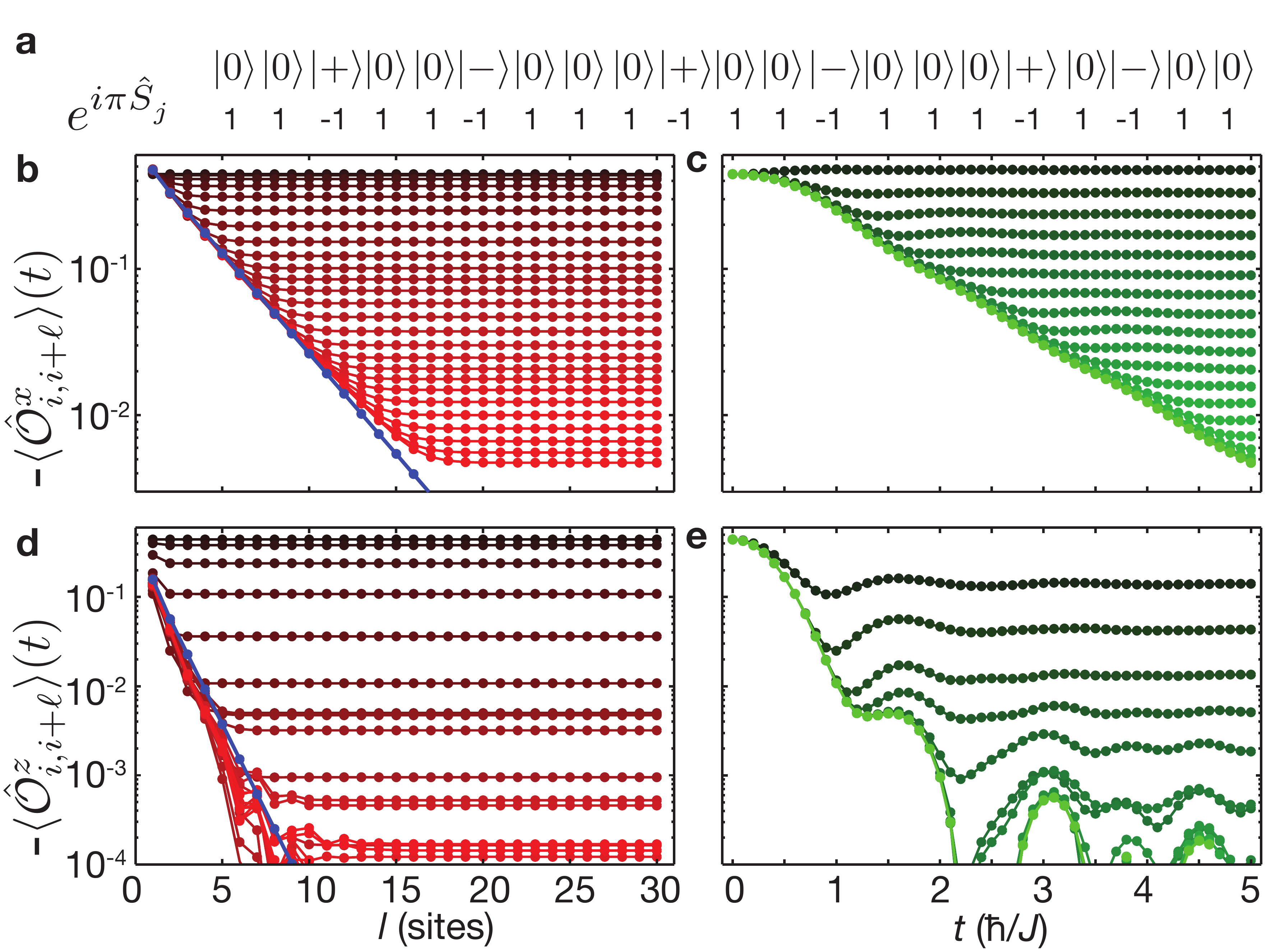}
  \caption{(color online) {\bf a:} 
    Sketch of a product state featuring the dilute 
    anti-ferromagnetic order detected by $\hat{\mathcal O}^z_{i,j}$. 
    {\bf b}-{\bf e:} String correlations for a quench of $\ket{\Psi_0}$ with 
    Hamiltonian~\eqref{eq:XXZ}, $\Delta = 0.2$. 
    {\bf b:} $\langle \hat{\mathcal O}^{x}_{i,i + \ell}\rangle (t)$ 
    as a function of the length $\ell$. Each line refers to different time, 
    ranging from $t = 0$~$\hbar/J$ (dark red) to $t = 5.0$~$\hbar /J$ (bright red), 
    with spacing $0.2$~$\hbar / J$. The blue line represents string correlations 
    of the thermal state $\hat \rho_{\rm th}(\beta)$, with $\beta = 1.6$ $J^{-1}$.
    {\bf c:} $\langle \hat{\mathcal O}^{x}_{i,i + \ell}\rangle (t)$ as a function of $t$. 
    Each line refers to different $\ell$, ranging from $\ell =1$ (dark green) 
    to $\ell = 30$ (bright green), with spacing $1$. 
    {\bf d}-{\bf e:} $\langle \hat{\mathcal O}^{z}_{i,i + \ell}\rangle (t)$ 
    with color coding and axes as in {\bf b}-{\bf c}, respectively.}
  \label{fig:Figure1}
\end{figure}

{\it Quantum quenches and string order.}---As a paradigm to study out-of-equilibrium SO,
we consider the Haldane phase in spin-1 infinite chains, a symmetry-protected 
topological phase~\cite{Pollman2010, Pollmann2012} characterized by SO:~\cite{denNijs1989} 
\begin{equation}
  \mathcal O^{\alpha} \doteqdot \lim_{|i-j| \to \infty} \langle \hat{\mathcal O}^\alpha_{i,j} \rangle \neq 0, 
  \label{eq:def:SO}
\end{equation}
where $ \hat{\mathcal O}^\alpha_{i,j} \doteqdot 
\hat S^{\alpha}_i \otimes e^{i \pi \hat S^\alpha_{i+1}} \ldots e^{i \pi \hat S^{\alpha}_{j-1}}
\otimes \hat S^{\alpha}_j$ and $\hbar \hat S^{\alpha}_{j}$ 
is the $\alpha=x,y,z$ component of the spin operator at the $j$-th site.
SO signals the presence of a ``dilute'' anti-ferromagnetic order where oriented spins 
$S^{\alpha}_i= \pm 1$ spatially alternate, although they can be separated by 
arbitrary-length strings of $S_i^{\alpha} =0$ spins (Fig.~\ref{fig:Figure1}~\textbf{a}).
Spin-1 chains can also effectively mimic the physics of cold bosonic gases 
trapped in 1D optical lattices, where the Haldane phase 
may appear.~\cite{DallaTorre2006, Rossini2012} 

The ground state of the Affleck-Kennedy-Lieb-Tasaki (AKLT) model in the $\sum_i \langle \hat S^z_i \rangle = 1$ 
magnetization sector, $\ket {\Psi_0}$, is taken as initial state.~\cite{AKLT} 
It belongs to the Haldane phase and admits an exact 
matrix-product-state (MPS) description with bond link~\cite{Kluemper1993} $D=2$, which 
lowers technical intricacies and allows the investigation of longer time evolutions. 
Next, $\ket {\Psi_0}$ is let free to evolve unitarily according to the
Hamiltonian $\hat{\Ham}$, such that $\ket {\Psi(t)} \doteqdot e^{-i \hat \Ham t / \hbar} \ket {\Psi_0}$. 
In the following, we characterize the SO dynamics by investigating time 
and length dependence of $\langle \hat {\mathcal O}_{i,i+\ell}^\alpha \rangle (t)$, 
where $\langle \cdot \rangle (t)$ denotes the expectation value over $\ket {\Psi (t)}$. 
We employ several numerical algorithms based on MPS:~\cite{Schollw2011,note} 
$\ket{\Psi (t)}$ is computed with a time-evolving block-decimation (TEBD) 
technique,~\cite{Vidal_2004} whereas thermal properties are obtained 
with the ancilla method.~\cite{Feiguin2005}

{\it String operators in spin-1 chains.}---We start considering the anti-ferromagnetic spin-1 XXZ model: 
\begin{equation}
  \hat{\Ham} = J \sum_i \left( \hat S_i^x \hat S_{i+1}^x + \hat S_i^y \hat S_{i+1}^y + 
  \Delta \hat S_i^z \hat S_{i+1}^z \right), 
  \label{eq:XXZ}
\end{equation}
with $J>0$. For simplicity, we assume $\Delta > 0$: 
in this case, when $\Delta < \Delta_c$ ($\Delta_c \sim 1.18$) the ground state 
is in the Haldane phase, whereas for $\Delta > \Delta_c$ it presents N\'eel order.~\cite{Chen2003} 
The model is integrable only in the classical limit $\Delta \to \infty$.

We first discuss the case $\Delta=0.2$ (Fig.~\ref{fig:Figure1}).
For any finite $t>0$, the state displays SO, because the expectation value of longer 
string operators saturates to a non-zero plateau as a function of length. 
The dynamics of the string operator in the {\it x}-direction, 
$\langle \hat {\mathcal O}_{i,i+\ell}^x \rangle (t)$, exhibits two regions separated 
by a length scale $L_{\rm th} (t)$. For longer strings, $\ell\gg L_{\rm th} (t)$, such operator 
is exponentially close to its infinite-length value: 
$ \langle \hat {\mathcal O}_{i,i+\ell}^x \rangle (t)\approx \mathcal O^x (t)$.
On the opposite, for shorter strings, $\ell \ll L_{\rm th} (t)$, it is close 
to its infinite-time value: $\langle \hat {\mathcal O}_{i,i+\ell}^x \rangle (t)
\approx \lim_{t \to \infty} \langle \hat {\mathcal O}_{i,i+\ell}^x \rangle (t )$. 
As we discuss below, the long-string region can be understood in terms of a LR bound, 
while short strings suggest the appearance of a thermal region growing with time. 
Data for the string in the {\it z}-direction, 
$\langle \hat {\mathcal O}_{i,i+\ell}^z \rangle (t)$, indicate a similar behavior, 
which is, however, less conclusive, due to superimposed damped 
oscillations in time. [Note that $\langle \hat {\mathcal O}_{i,i+\ell}^z \rangle (t)$ 
is typically an order of magnitude smaller than $\langle \hat {\mathcal O}_{i,i+\ell}^x \rangle (t)$].

The long-string behavior is best understood via the following key observation. 
For Hamiltonian~\eqref{eq:XXZ}, string correlations satisfy a LR bound:~\cite{Lieb1972, Bravyi2006}
\begin{equation}
  \big| \langle \hat{ \mathcal O}^{\alpha}_{i,i+\ell}\rangle (t) - \mathcal O^{\alpha} (t) \big| 
  < C e^{- (\ell - vt) / \xi }.
  \label{eq:LR:bound}
\end{equation}
Equation~\eqref{eq:LR:bound} follows directly from the Kennedy-Tasaki 
transformation,~\cite{Kennedy1992, Oshikawa1992} 
$\hat {\cal U}_{\rm KT} = \prod_{j<k} \exp (i \pi \hat S^z_j \hat S^x_k)$, which is unitary 
and non-local. This dual transformation maps~\eqref{eq:XXZ} 
into another local model, which is ferromagnetic and for $0 \leq \Delta \leq \Delta_c$ 
has four symmetry-broken magnetic ground states.
String operators are mapped to two-point operators: 
$\hat {\cal U}_{\rm KT} \hat {\mathcal O}^\alpha_{i,j} \hat {\cal U}_{\rm KT}^\dagger = 
\hat S^\alpha_i \hat S^\alpha_j $. 
In this dual picture, spin-spin correlations obey the LR bound 
$\big| \langle \hat{ S}^{\alpha}_{i} \hat{ S}^{\alpha}_{i+\ell}\rangle (t) 
- \langle \hat{ S}^{\alpha}_{i} \rangle \langle \hat{ S}^{\alpha}_{i+\ell}\rangle (t) \big| 
< C e^{- (\ell - vt) / \xi}$.~\cite{Bravyi2006} 
Indeed, take as the initial state $\hat {\cal U}^\dagger_{\rm KT} \ket \Psi$: 
the insertion of $\hat {\cal U}_{\rm KT} \hat {\cal U}^\dagger_{\rm KT}$ in every 
expectation value and the fact that 
$ \mathcal O^{\alpha} \doteqdot \lim_{|i-j| \to \infty} \langle \hat{\mathcal O}^\alpha_{i,j} \rangle 
= \langle \hat {\cal U}_{\rm KT}^\dagger \hat S^\alpha_i \hat {\cal U}_{\rm KT} \rangle^2$ 
yield Eq.~\eqref{eq:LR:bound}. In general, a LR bound for string operators 
can be derived whenever there is a mapping to a dual model such that 
the string operator is mapped to a two-site correlator obeying a LR 
bound.~\cite{Calabrese2011}

\begin{figure}
  \includegraphics[width=\columnwidth]{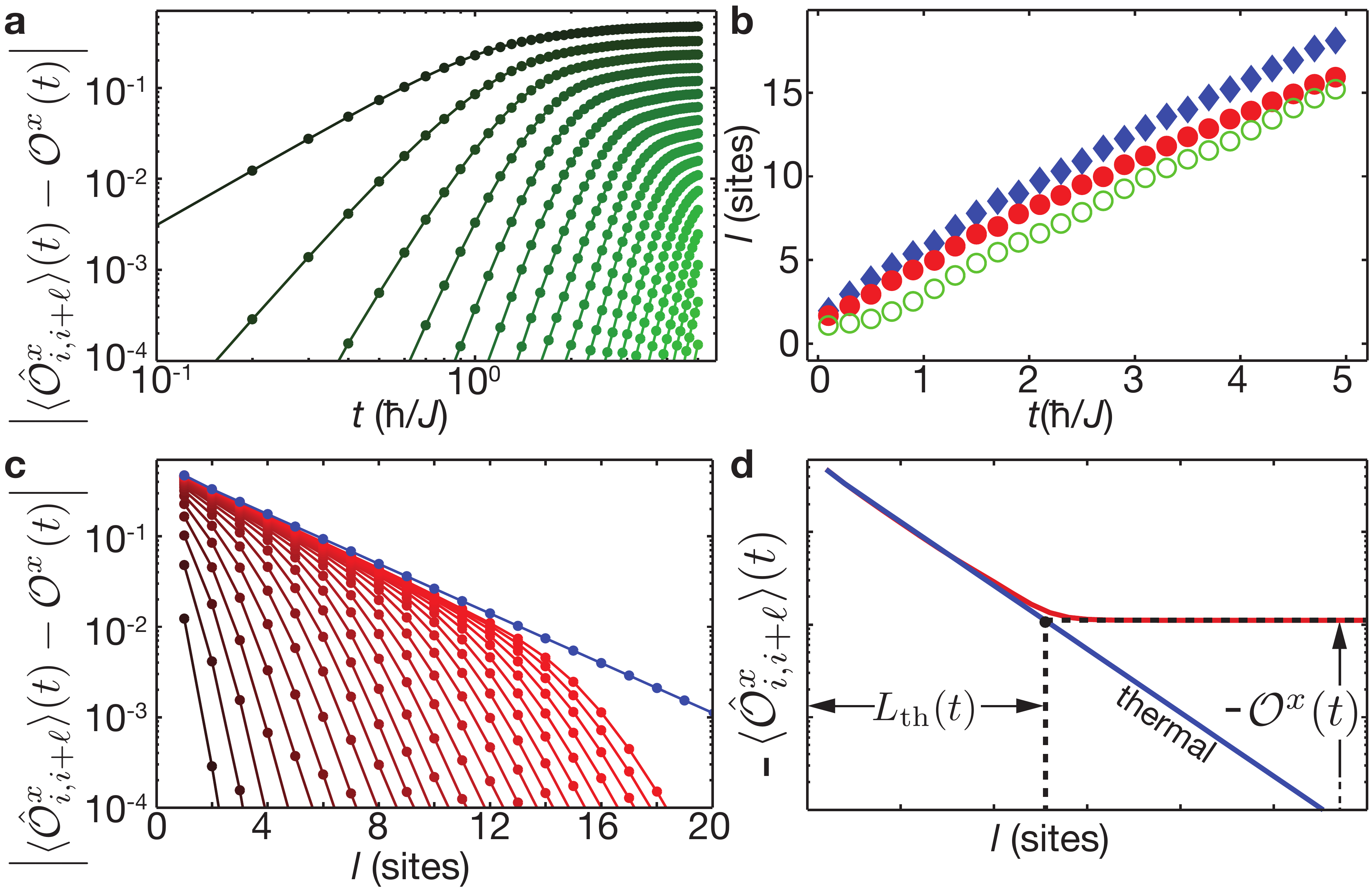}
  \caption{(color online) 
    String correlations for a quench with Hamiltonian~\eqref{eq:XXZ}, $\Delta = 0.2$. 
    {\bf a:} Subtracted string correlations 
    $|\langle \hat{\mathcal O}^{x}_{i,i + \ell}\rangle (t) - \mathcal O^{x}(t)|$ 
    as a function of time. Color code as in Fig.~\ref{fig:Figure1}{\bf c}. 
    {\bf b:} Spreading of string correlations. 
    Space-time dependence of $t_{F_0} (\ell)$ for $F_0 = 10^{-4}$ (blue diamonds), 
    $10^{-3}$ (red dots). 
    Time dependence of $L_{\rm th}(t)$ (green empty circles). 
    {\bf c:} Subtracted string correlations 
    $|\langle \hat{\mathcal O}^{x}_{i,i + \ell}\rangle (t) - \mathcal O^{x}(t)|$ 
    as a function of length. Color code as in Fig.\,\ref{fig:Figure1}{\bf b}. 
    The blue line represents string correlations of the thermal state 
    $\hat \rho_{\rm th}(\beta)$, with $\beta = 1.6$ $J^{-1}$. 
    {\bf d:} Illustration of the construction of $L_{\rm th}(t)$.}
  \label{fig:SO:LR}
\end{figure}

In Fig.~\ref{fig:SO:LR}, we study the subtracted correlations of Eq.~\eqref{eq:LR:bound}. 
A causal-cone effect is observed. Correlations are initially zero and increase 
more slowly the longer the considered string is (panel {\bf a}). 
The dispersion relation of the cone can be extracted by identifying 
the time $t_{F_0}(\ell)$ satisfying 
$|\langle \mathcal O_{i,i+\ell}^x \rangle (t) - \mathcal O^\alpha(t)| = F_0$.
A linear space-time dependence appears, $t_{F_0}(\ell) \sim \ell / v_{\rm LR}$, 
signaling a ballistic spreading of string correlations for short times, 
even if the system is non-integrable (panel {\bf b}).
Panel {\bf c} points out that the value $\mathcal O^x$ is approached 
exponentially as a function of length, as predicted in Eq.~\eqref{eq:LR:bound}.

For short strings, the behavior of 
$ \langle \hat {\mathcal O}^x_{i,i+\ell} \rangle (t)$ appears to be well captured by a thermal 
ensemble,~\cite{note1} 
$\hat \rho_{\rm th}(\beta) = e^{- \beta \hat H} /\mathcal Z$ ($\mathcal Z$ is the partition 
function), whose effective temperature, $\beta$, is defined by
$\bra{\Psi_0} \hat {\Ham} \ket{\Psi_0} = \text{tr} [\hat {\Ham} \, \hat \rho_{\rm th} (\beta) ]$.
As shown in Fig.~\ref{fig:Figure1}, the SO of $\ket{\Psi (t)}$ can be characterized 
in this way for strings up to length $\ell = 10$. 
Note that $\hat {\mathcal O}^\alpha_{i, i+\ell}$ probes the properties of a multi-site region, 
thus providing information not included in any two-point observable. 
The typical length scale, $L_{\rm th} (t)$, separating the thermal  
from the LR region, can be defined according to 
$\tr [\hat {\mathcal O}^{x}_{i,i+L_{\rm th}(t)} \, \hat \rho_{\rm th}(\beta)] = \mathcal O^{x}(t)$, 
where $L_{\rm th} (t)$ is the length at which the thermal value of the string 
operator crosses the infinite-length value of the dynamically evolved string operator 
(see Fig.~\ref{fig:SO:LR}~{\bf d}). 
Figure~\ref{fig:SO:LR}~{\bf b} shows a linear space-time dependence $L_{\rm th} (t) \sim v_{\rm th} t$, 
compatible to $v_{\rm LR} t_{F_0}$.

\begin{figure}[t]
  \includegraphics[width=\columnwidth]{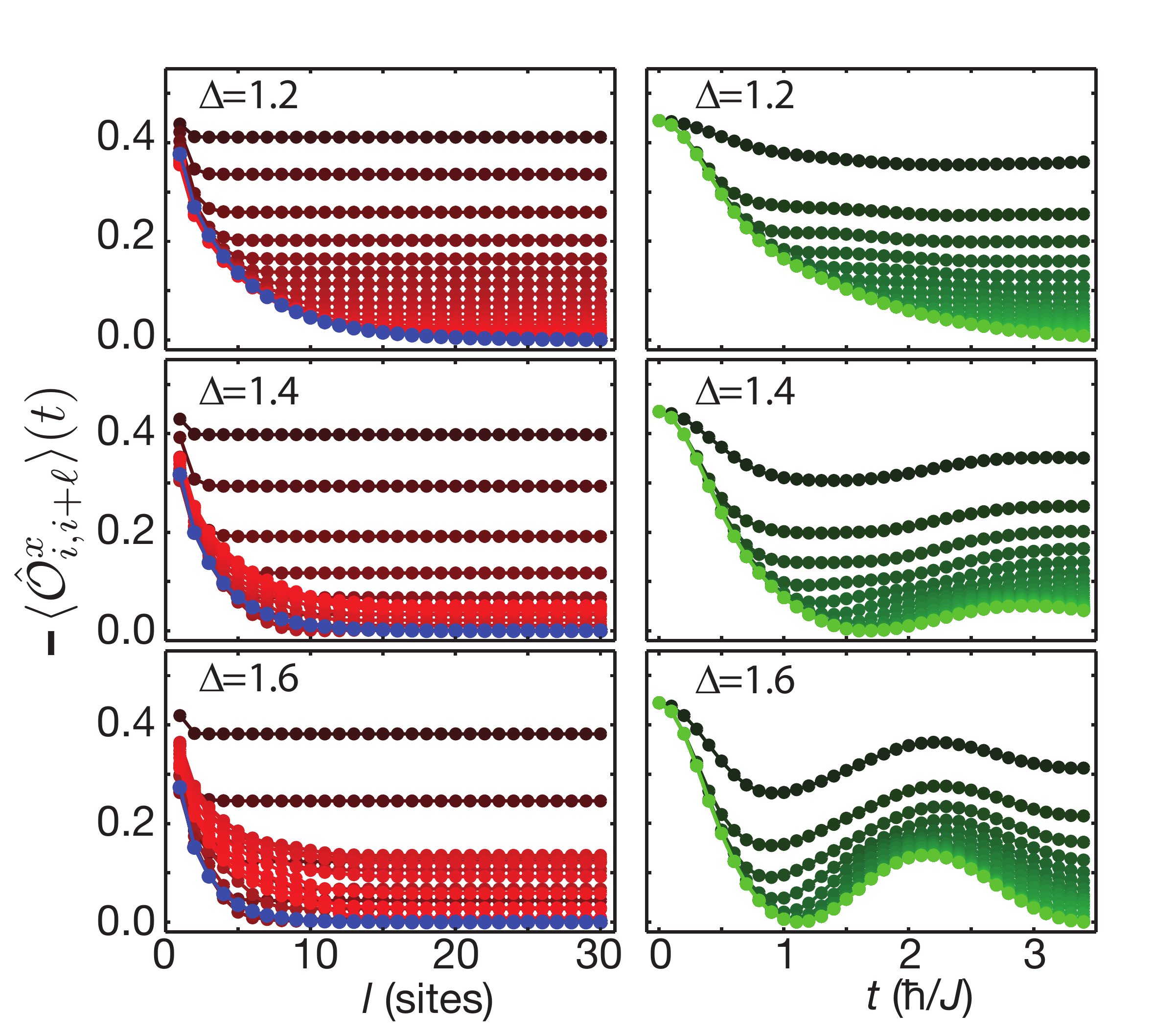}
  \caption{(color online) 
    String correlations $\langle \mathcal O^{x}_{i,i + \ell}\rangle (t)$ for a quench 
    with Hamiltonian~\eqref{eq:XXZ} for $\Delta=1.2,1.4$ and $1.6$,
    as a function of length $\ell$ (left panels), and time $t$ (right panels). 
    Left panels: each line refers to a different time, 
    ranging from $t = 0$~$\hbar/J$ (dark red) to $t = 3.4$~$\hbar /J$ (bright red), 
    with spacing $0.2$~$\hbar / J$. 
    Blue lines represent string correlations of the thermal states $\hat \rho_{\rm th}(\beta)$, 
    with $\beta = 1.92$, $1.48$ and $1.2$ $J^{-1}$ respectively.
    Right panels: each line refers to a different length, ranging from $\ell =1$ (dark green) 
    to $\ell = 30$ (bright green), with spacing $1$.}
  \label{fig:figure3}
\end{figure}

We now extend the previous analysis to other values of $\Delta$. 
The presence of an increasing region with size $v t$, where string correlations 
$\langle \hat{\mathcal O}^x_{i,j} \rangle (t)$ are thermal-like, approximately holds 
for $\Delta \lesssim 1.2$ (see Fig.~\ref{fig:figure3}). 
For larger values of $\Delta$, an oscillatory behavior sets in, 
complicating the short-time evolution. 
From our data, it is not possible to tell whether this is due 
to approaching the integrable point $\Delta \to +\infty$, or 
the phase transition at $\Delta_c\sim1.18$. 
Note that for all our cases $\beta / E_{\rm gap} < 10$, where
$E_{\rm gap}$ is the second energy gap for $L=60$ computed for
$\sum_i \langle \hat S^z_i \rangle = 0$. (Two degenerate ground states are
expected in the thermodynamic limit.)

A natural explanation of our data goes as follows: 
the density matrix $\hat \rho_{\mathrm S} = \text{tr}_{\rm chain / S} 
\left[\ket {\Psi(t)} \bra {\Psi(t)} \right]$ of a region $S$ of size $vt$ appears to be thermal.
This interpretation is corroborated by our study of $\hat \rho_S$ for regions 
up to three sites and by the study of $\langle \hat S^x_i \hat S^x_j \rangle (t)$ 
correlations (not shown). 
The behavior of $\langle \hat {\mathcal O}^{z}_{i,i+\ell} \rangle (t)$ does not fit 
immediately in this framework, see Fig.~\ref{fig:Figure1}~\textbf{d},~\textbf{e}. 
Even if short strings equilibrate (with superimposed damped oscillations) 
within accessible times, this happens more slowly than for 
$\langle \hat {\mathcal O}^{x}_{i,i+\ell} \rangle (t)$, hinting to the possibility 
that different internal degrees of freedom equilibrate in significantly different ways.

{\it Symmetry protection.}---In the previous discussion we stressed
that $\ket{\Psi (t)}$ exhibits SO for every accessible $t$. 
This is not always the case and depends on the symmetries of $\hat {\Ham}$. 
We now prove a necessary and sufficient condition on $\hat {\Ham}$ 
for $\ket{\Psi (t)}$ to display SO at short times.

Let us recall that a \textit{necessary} condition for a state $\ket \Phi$ to possess 
the SO of Eq.~\eqref{eq:def:SO} is to be symmetric under the action 
of a global $e^{i \pi \sum_j \hat S^\alpha_j}$ transformation: 
$e^{i \pi \sum_j \hat S^\alpha_j} \ket \Phi = e^{i \theta_\alpha} \ket \Phi$.~\cite{PerezGarcia2008} 
We first investigate necessary and sufficient conditions for:
$e^{i \pi \sum_j \hat S^\alpha_j} \ket {\Psi(t)} = \ket {\Psi(t)}$, $\forall t$.
We focus on the Abelian ${\mathbb D}_2$ group of $\pi$ rotations 
along three orthogonal axes $e ^{i \pi \sum_j \hat S^{\alpha}_j}$, $\alpha = x,y,z$.
Note that $e^{i \theta_\alpha} =1$ $\forall t$ for $\alpha =x,z$ because
$ e^{i \pi \sum_j \hat S^\alpha_j} \ket {\Psi_0} = \ket {\Psi_0}$, the time evolution 
is continuous and the eigenvalues of $e^{i \pi \sum_j \hat S^\alpha_j}$ are discrete.

We begin considering an infinitesimal time $\delta t$, 
so that $e^{- i \hat \Ham \delta t / \hbar} \approx \mathbb I -i (\delta t / \hbar) \hat \Ham$. 
Let us decompose $\hat \Ham = \hat \Ham_{\mathrm e} + \hat \Ham_{\mathrm o}$, 
$\hat \Ham_{\mathrm e}$ being the symmetric part of the Hamiltonian: 
$e^{i \pi \sum_j \hat S^\alpha_j} \hat \Ham_{\mathrm e} e^{-i \pi \sum_j \hat S^\alpha_j} = \hat \Ham_{\mathrm e}$ 
for $\alpha = x,z$, and $\hat \Ham_{\mathrm o}$ the rest. 
Clearly, when the Hamiltonian is symmetric ($\hat \Ham_{\mathrm o} = 0$), 
$\ket {\Psi (t)}$ is $\mathbb D_2$ symmetric.
This is true also in presence of non-symmetric contributions 
($\hat \Ham_{\mathrm o} \neq 0$). At first order in $\delta t$, this is equivalent 
to $\hat \Ham_{\mathrm o} \ket{\Psi_0} = 0$ because $\hat \Ham_{\mathrm o}$ connects $\ket {\Psi_0}$ 
only to states which are not left invariant by the action of every element of the ${\mathbb D}_2$ group.
A full series expansion of $e^{-i \hat{\mathcal H} t /\hbar}$ yields 
that the $\mathbb D_2$ symmetry of the state $\ket{\Psi (t)}$ 
is equivalent to:
\begin{equation}
  \hat \Ham_{\mathrm o} \big( \hat \Ham_{\mathrm e} \big)^n \ket{\Psi_0} = 0, 
  \quad \forall n \in \mathbb N_0.
  \label{eq:main:condition}
\end{equation}
Since $\mathbb D_2$ symmetry is necessary for SO,~\eqref{eq:main:condition} 
is a necessary condition for $\ket{\Psi (t)}$ to display SO.

To show that it is also sufficient, we prove continuity of $\mathcal O^\alpha (t)$ 
as a function of $t$ ($\mathcal O^\alpha \neq 0$ at $t=0$).
Let us observe that $ \langle \hat {\mathcal O}^\alpha_{i,j} \rangle (\delta t) = 
\langle \hat {\mathcal O}_{i,j}^\alpha \rangle (0) 
+ i (\delta t /\hbar) \bra{ \Psi_0} [\hat { \mathcal H_{\rm e}}, 
\hat {\mathcal O}^\alpha_{i,j}] \ket{\Psi_0} + o (\delta t^2)$ 
where $\hat {\mathcal H}_{\rm o}$ is discarded because of~\eqref{eq:main:condition}.
We assume $\hat{\mathcal H}_{\rm e}$ to be short ranged: 
$ \hat{\mathcal H}_{\rm e} = \sum_j \hat h_{\mathrm e,j}$, each $\hat h_{\mathrm e, j}$ 
having support on $k$ neighboring spins. 
Thus the operator $ [\hat { \mathcal H_{\rm e}}, \hat {\mathcal O}^\alpha_{i,j}]$ 
has support at most on $2k$ spins centered around $i$ and around $j$, 
and is bounded even for $|i-j|\to \infty$.
Using~\eqref{eq:main:condition} one can prove that terms
of order $(\delta t/\hbar)^{n}$ for $n > 1$ are bounded as well.
Continuity follows, so that Eq.~\eqref{eq:main:condition} is equivalent 
to the presence of SO at finite times. 

Remarkably, SO is more resilient to perturbations in a dynamical context than in a static one. 
Indeed, a necessary condition for the ground state of $\hat {\mathcal H}$ to display SO 
in Eq.~\eqref{eq:def:SO} is that $\hat {\mathcal H}$ commutes with $ e^{i \pi \sum_j \hat S^\alpha_j} $.~\cite{Pollman2010} 
However, SO can withstand a quench by a non-symmetric Hamiltonian, 
as long as~\eqref{eq:main:condition} holds.
An example is the spin-1 XXZ model~\eqref{eq:XXZ} with
an additional magnetic field along {\it z}. 
Eq.~\eqref{eq:main:condition} enforces that $\ket {\Psi (t)}$ is
in the kernel of $\hat \Ham_{\mathrm o}$, so that effectively it
can be computed only considering $\hat \Ham_{\mathrm e}$.

\begin{figure}
  \includegraphics[width=\columnwidth]{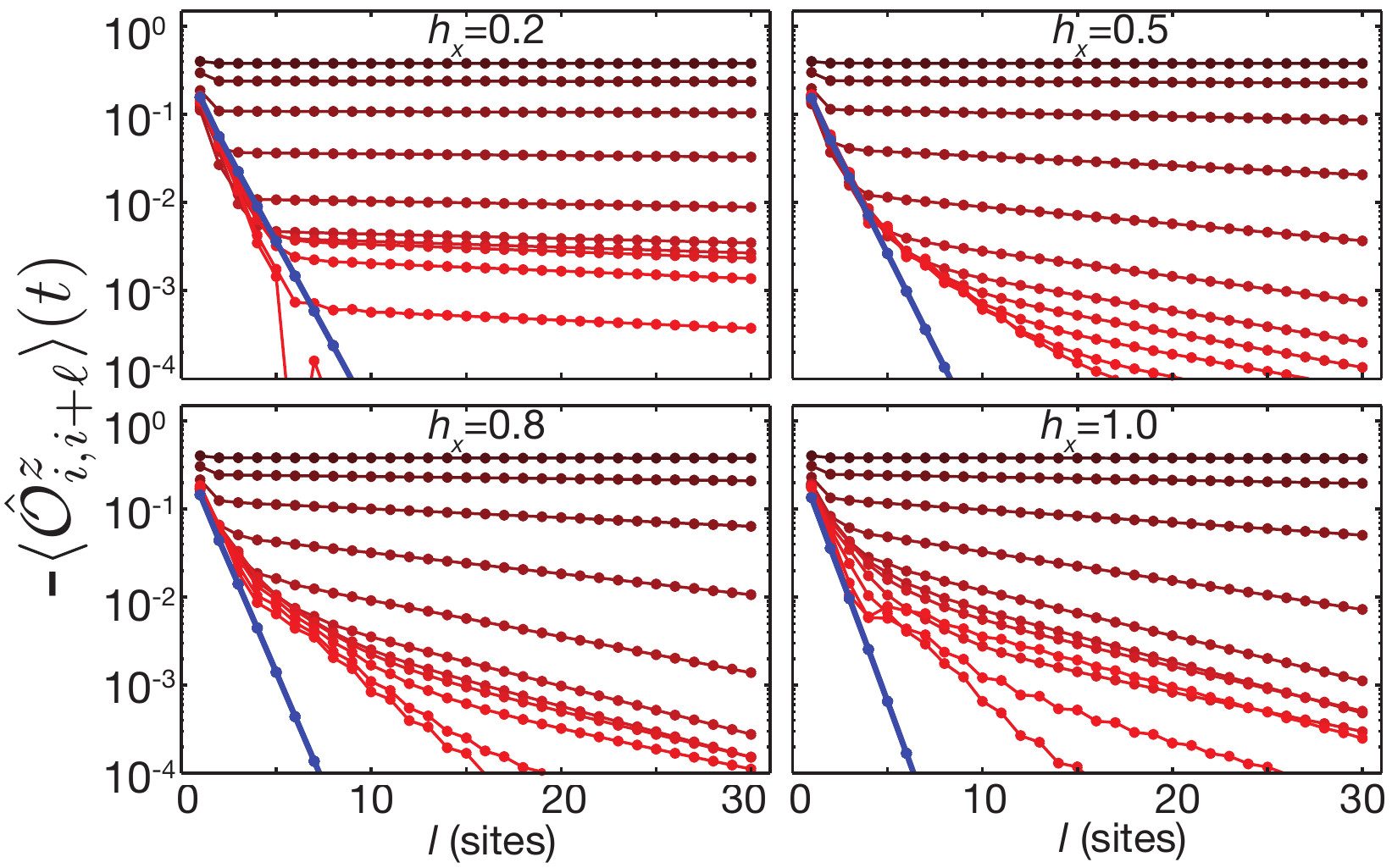}
  \caption{(color online) 
    String correlations $\langle \hat{\mathcal O}^{z}_{i,i + \ell}\rangle (t)$ 
    as a function of $\ell$ for a quench with Hamiltonian~\eqref{eq:XXZ:Bx}, 
    with $\Delta = 0.2$, $h_x = 0.2$, $0.5$, $0.8$ and $1.0$.
    Each line refers to a different time, ranging from $t = 0$~$\hbar/J$ (dark red) 
    to $t = 2.2$~$\hbar/J$ (bright red) with spacing $0.2$~$\hbar/J$.
    Blue lines represent string correlations of the thermal states,
    $\hat \rho_{\rm th}(\beta)$, with $\beta = 1.56$, $1.52$, $1.44$ and $1.32$ $J^{-1}$ respectively.}
  \label{fig:Figure4b}
\end{figure}

Symmetric Hamiltonians $\hat \Ham_{\rm e}$, such as~\eqref{eq:XXZ}, are related to spin-1 local models 
via the duality mapping $\hat {\mathcal U}_{\rm KT}$~\cite{Pollman2010}.
In this dual formulation, string correlations are mapped to two-point correlations.
It is interesting to consider models that do not satisfy~\eqref{eq:main:condition} 
and thus such that, already after an infinitesimal time, SO is lost: 
$\lim_{\ell \to \infty} \langle \hat{\mathcal O}^\alpha_{i,i+\ell} \rangle (t) = 0$, $\forall t>0$. 
We consider a spin-1 XXZ model in a transverse field:
\begin{equation}
  \mathcal {\hat H}_x \! = \!
  J \sum_i \left( \!
  \hat S_i^x \hat S_{i+1}^x + \hat S_i^y \hat S_{i+1}^y + \Delta \hat S_i^z \hat S_{i+1}^z 
  + h_x \hat S^{x}_i \! \right) . 
  \label{eq:XXZ:Bx}
\end{equation}
Since the model has a $\mathbb Z_2$-symmetry of $\pi$ rotations around $\hat x$, 
it still displays SO, and $\mathcal O^x \neq 0$, but $\mathcal O^z, \mathcal O^y = 0$.

In Fig.~\ref{fig:Figure4b}, we plot the behavior of string correlations 
$\langle \hat {\mathcal O}^z_{i,j} \rangle (t)$ for $\Delta = 0.2$ and several values 
of $h_x$: as expected, for $h_x \neq 0$, they do not saturate but rather decay to zero 
as a function of length. 
Interestingly, a common decay curve for short strings 
appears, from which correlations depart when $\ell$ overcomes a given length scale increasing in time. 
This behavior is qualitatively similar 
to that for the spin-1 XXZ model (see Fig.~\ref{fig:Figure1}). 
Again, the equilibrium properties 
of small regions can be related to those of a thermal density matrix.

\begin{figure}[!t]
  \includegraphics[width=\columnwidth]{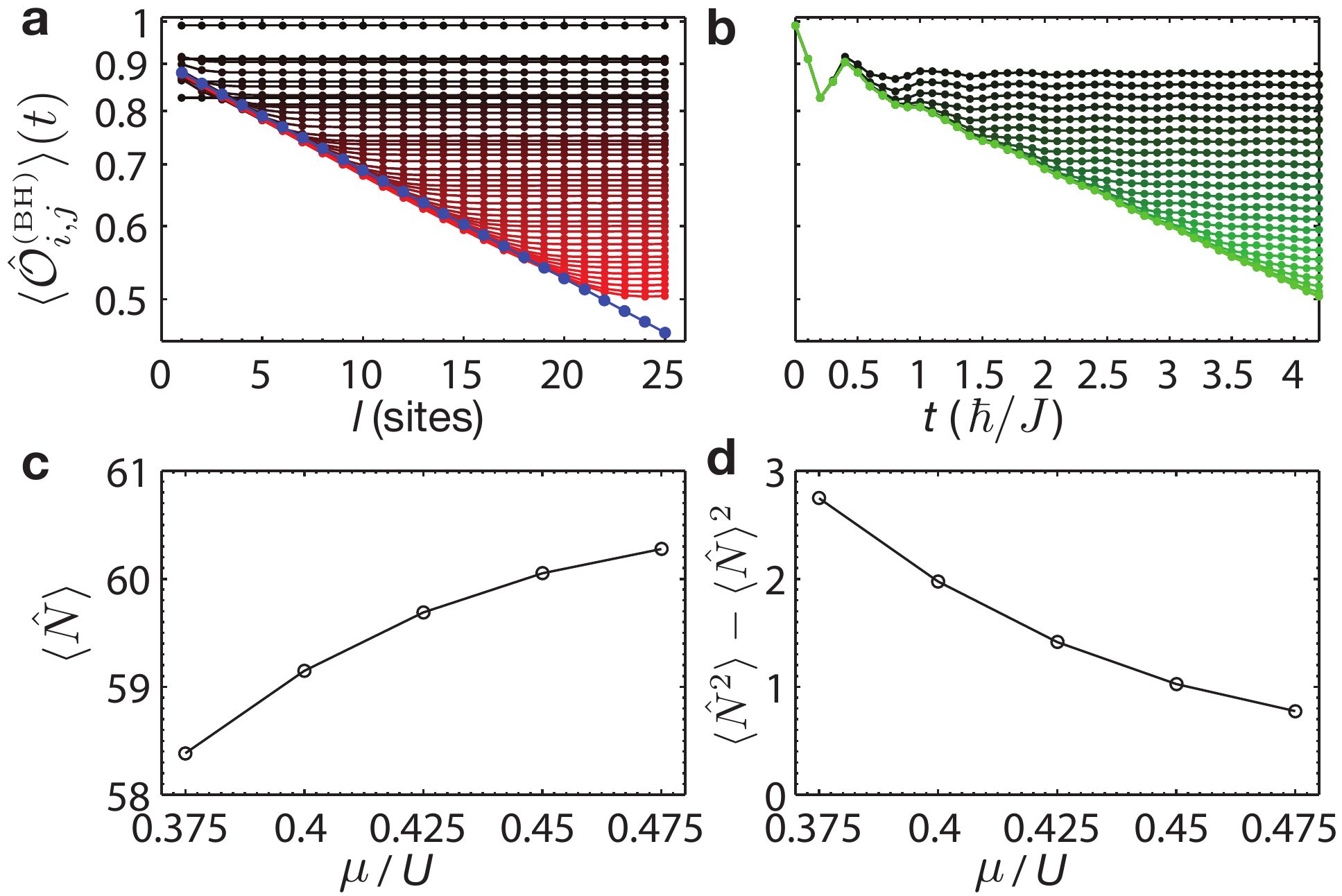}
  \caption{(color online) 
    String correlations $\langle \hat{\mathcal O}^{\rm \scriptscriptstyle (BH)}_{i,i + \ell}\rangle (t)$ 
    as a function of $\ell$ for a quench with Hamiltonian~\eqref{eq:BH}, 
    from $U_0 = 40 J$ to $U = 15 J$.
    {\bf a:} Each line refers to a different time, ranging from $t = 0$~$\hbar/J$ (dark red) 
    to $t = 4.2$~$\hbar/J$ (bright red) with spacing $0.1$~$\hbar/J$.
    The blue line represents string correlations of the grand canonical state 
    $\hat \rho_{\rm gc}(\beta, \mu)$, with $\beta = 0.788 J^{-1}$, and $\mu = 0.475 U_{\rm Q}$. 
    {\bf b:} Each line refers to a different length, ranging from $\ell =1$ (dark green) 
    to $\ell = 26$ (bright green), with spacing $1$.
    {\bf c}-{\bf d:} Mean and variance of the particle number
    for $\hat \rho_{\rm gc}(\beta, \mu)$ as a function of $\mu$, for the $\beta$ 
    matching the initial state energy: 
    $0.578$, $0.622$, $0.67$, $0.724$ and $0.786 J^{-1}$ respectively.}
  \label{fig:Figure4}
\end{figure}

{\it String operators in the Bose-Hubbard model.}---The out-of-equilibrium study 
of string correlations is within the immediate experimental reach.
Ultracold bosonic atoms prepared in 1D tubes with a deep optical lattice 
along the system direction can be described by a BH model
\begin{equation}
  \hat \Ham_{\rm BH} = - J \sum_j ( \hat a^\dagger_j \hat a_{j+1} + {\rm H.c.}) 
  + \frac{U}{2} \sum_j \hat n_j (\hat n_j-1) \,,
  \label{eq:BH}
\end{equation}
where $\hat a^\dagger_j$ creates one boson on site $j$, $\hat n_j = \hat a^\dagger_j \hat a_j$,
$J$ is the hopping rate and $U$ the interaction strength.
For $U \gg J$ the model displays a Mott insulating (MI) phase characterized by the string observable 
$\hat{\mathcal O}^{\rm \scriptscriptstyle (BH)}_{i,j} \doteqdot e^{i \pi \delta \hat{n}_{i}} 
\otimes e^{i \pi \delta \hat n_{i+1}} \ldots e^{i \pi \delta \hat n_{j-1}} \otimes e^{i \pi \delta\hat n_{j}}$, 
$\delta \hat n_j$ being
the fluctuation over the average density of bosons at site $j$. 
String correlations for up to $9$ sites have been detected 
using a quantum gas microscope.~\cite{Endres_2011} 
Moreover, Ref.~\onlinecite{Cheneau_2012} reports an experimental study of the non-equilibrium dynamics 
of two-point correlations past a sudden quench with the same setup: 
starting from a deep MI at $U_0 = 40 J$, the lattice strength is lowered to
values of $U_{\rm Q}$ ranging between $5 J$ and $9 J$;
time scales of up to $2.6$ $\hbar /J$ have been reached.

The $\langle \hat{\mathcal O}^{\rm \scriptscriptstyle (BH)}_{i,j} \rangle(t)$ string propagation 
displays the same qualitative behavior of the spin-1 case
(see Fig.~\ref{fig:Figure4}, upper panels, for a typical situation).
The interpretation of long-time properties in terms of a grand canonical ensemble
$\hat \rho_{\rm gc}(\beta, \mu ) \doteqdot e^{-\beta (\hat \Ham_{\rm \scriptscriptstyle BH} - \mu \hat N)}/\mathcal{Z}'$
$(\hat N \doteqdot \sum_j \hat n_j)$ requires a careful analysis.
We identify a region $\Omega$ in the $(\beta,\mu)$ plane
where energy and number of particles of $\hat \rho_{\rm gc}(\beta, \mu)$
match those of the initial MI up to $10^{-3}$ accuracy.
The string properties of states inside $\Omega$ differ significantly,
mainly because $\Omega$ comprises a large range of values for $\beta$.
The lower panels of Fig.~\ref{fig:Figure4} show $\langle \hat N \rangle$
and $\langle \hat N^2 \rangle - \langle \hat N \rangle^2 $ for several states in the region.
The accuracy with which $\hat \rho_{\rm gc}(\beta,\mu)$ 
reproduces the long-time dynamics is non-trivially related to those expectation values,
so that the employment of a thermal state with fixed number of particles
could be necessary to describe the dynamical data.~\cite{note2}
Interestingly, it is possible to find an optimal
value $(\beta^*, \mu^*) \in \Omega$ reproducing 
$\langle \hat {\mathcal O}^{\rm (BH)}_{i,i+\ell}\rangle (t \to \infty)$.

Finally, we address the specific physics of the bosonic 
Haldane insulator,~\cite{DallaTorre2006, Rossini2012} 
the density-sector analogue of the Haldane phase in integer spin systems.
At present, two experimental challenges prevent its observation in a cold atomic gas:
(i) the engineering of nearest-neighbor interactions; 
(ii) the measurement of the appropriate string operator: 
$\hat{\mathcal O}^{\rm \scriptscriptstyle(HI)}_{i,j} \doteqdot \delta \hat n_i 
\otimes e^{i \pi \delta \hat n_{i+1}} \ldots e^{i \pi \delta \hat n_{j-1}} \otimes \delta \hat n_j$ 
(the operator $\hat n_i$ has not been accessed yet). 
Long-range interactions can be realized via dressed Rydberg gases, polar molecules 
or even atoms with permanent magnetic dipole. 
The on-site occupation number could be detected by preparing a single 1D system 
and letting the atoms tunnel orthogonally before the detection.
Finally, the presence of non-local order in other realistic cold-atom setups 
has been pointed out,~\cite{Montorsi2012, Rath2013} which 
may provide novel observational platforms.

{\it Conclusions.}---We have presented a detailed study of
the time-evolution of a string operator past a sudden quantum quench.
Such correlations signal a time-propagating horizon 
separating thermal from non-thermal regions.
The presence of a duality mapping allowed for
the formulation of a LR bound and for the 
interpretation of the data in terms of traveling quasiparticles.
The investigation of thermalization properties for Hamiltonians 
where duality mappings cannot be established
(which in our case do destroy SO abruptly and do not allow
for the formulation of a LR bound)
is an interesting problem left for a forthcoming publication.
Our results can be tested experimentally in one-dimensional
optical lattices in the Mott phase.

{\it Acknowledgements.}---We thank P. Calabrese and V. Giovannetti for invaluable support. 
We also acknowledge enlightening discussions with E. Altman, 
M. C. Ba\~{n}uls, M. Roncaglia, A. Silva and L. Tagliacozzo. 
This work was supported by EU (IP-SIQS), 
by Italian MIUR via PRIN Project 2010LLKJBX and via FIRB Project RBFR12NLNA, 
and by Regione Toscana POR FSE 2007-2013. 
M.E. thanks Scuola Normale Superiore for hospitality.


\begin{thebibliography}{99}
\bibitem{note1}
  Strictly speaking, because of the conservation of the total magnetization 
  $\hat{M}^z = \sum_i \hat S^z_i$, one has to deal with a generalized Gibbs ensemble
  $e^{-\beta \hat{\Ham} - \lambda \hat{M}^z} / \mathcal{Z}'$.
  However, since Hamiltonian~\eqref{eq:XXZ} has particle-hole symmetry, $\lambda = 0$
  enforces the constraint $\langle \hat{M}^z \rangle = 0$.
\end{thebibliography}
\end{document}